\documentclass[12pt]{iopart}
\usepackage{epsfig}
\usepackage{graphicx}
\usepackage{color}

\begin{document}
\title{Space-time Evolution of $J/\psi$ Production in High Energy Nuclear Collisions }
\author{Yunpeng Liu${}^{1}$, Zhen Qu${}^{1}$, Nu Xu${}^{2}$ and Pengfei Zhuang${}^{1}$}
\address{
${}^1$~Physics Department, Tsinghua University, Beijing 100084, China\\
${}^2$~Nuclear Science Division, Lawrence Berkeley National Laboratory, Berkeley, CA 94720, USA
}
\ead{liuyp06@mails.tsinghua.edu.cn}

\begin{abstract}
The space-time evolution of $J/\psi$ production in central Au+Au
collisions at RHIC energy is investigated in a transport model.
Both gluon dissociation and continuous regeneration of $J/\psi$s
inside the deconfined state are considered.
\end{abstract}
%\maketitle

$J/\psi$\ suppression~\cite{matsui} is widely accepted as an
essential signal of the quark-gluon plasma (QGP) formed in
relativistic heavy ion collisions. However, the mechanism of
$J/\psi$ production in hot and dense medium is still an open
question. Different from the $J/\psi$ suppression observed at
SPS~\cite{na50} where almost all the charmonia are produced in the
initial stage via hard processes and then suffer from the
anomalous suppression in the QGP, there is a remarkable number of
charm quarks in the QGP produced at RHIC energy and the
recombination of those uncorrelated charm quarks offers another
origin of $J/\psi$ production~\cite{thews1}. Both sudden
generation on the hadronization surface in grand canonical
ensemble~\cite{pbm1} or canonical ensemble~\cite{gorenstein} and
continuous regeneration~\cite{rapp1,thews2} inside the QGP are
discussed to describe the charmonia production.

Since the charmonia are so heavy, they are difficult to be fully
thermalized in the QGP with light quarks and gluons as
constitutes, and a natural way to describe the charmonium motion
in hot and dense medium is through a transport approach. In this
paper, we investigate the space-time evolution of charmonium
production and calculate the nuclear modification factor $R_{AA}$
and averaged transverse momentum square $\langle p_t^2\rangle$ for
$J/\psi$s at RHIC energy in a transport model~\cite{zhuang}, where
the QGP is characterized by hydrodynamic equations and the
charmonium motion is controlled by a classical transport equation.
Both initial production and regeneration of charmonia and both
nuclear absorption and anomalous suppression can be
self-consistently considered in the model. The leakage
effect~\cite{matsui, blaizot, karsch1}, which is especially
important for describing the transverse momentum saturation at SPS
energy~\cite{hufner1}, is reflected in the free streaming term of
the transport equation.

Since a charmonium mass is much larger than the typical
temperature of the fireball created at RHIC, it is a good
approximation to describe the charmonium distribution function
$f_\Psi({\bf p}_t,{\bf x}_t,\tau|{\bf b})$ for $\Psi=J/\psi,
\psi', \chi_c$ in the transverse phase space $({\bf p}_t,{\bf
x}_t)$ at time $\tau$ and fixed impact parameter ${\bf b}$ by a
Boltzmann equation~\cite{zhuang},
\begin{equation}
\label{trans}
\partial f_\Psi/\partial \tau +{\bf
v}_\Psi\cdot{\bf \nabla}f_\Psi = -\alpha_\Psi f_\Psi +\beta_\Psi.
\end{equation}
The second term on the left hand side arises from the
free-streaming of $\Psi$ with transverse velocity ${\bf v}_\Psi =
{\bf p}_t/\sqrt{{\bf p}_t^2+m_\Psi ^2}$, which leads to the
leakage effect and is important to high momentum
charmonia~\cite{hufner1}. The first and second terms on the right
hand side are respectively the suppression and regeneration rates
of charmonia. The former is usually taken from the gluon
dissociation process $J/\psi+g\rightarrow
c+\bar{c}$~\cite{peskin}. For $J/\psi$\ the cross section reads
\begin{eqnarray}
 \sigma_{J/\psi}(\omega)=A_0\frac{(\omega/\epsilon_{J/\psi}-1)^{3/2}}{(\omega/\epsilon_{J/\psi})^5},
\label{sigma}
\end{eqnarray}
where $\omega$\ is the gluon energy relative to $J/\psi$, and
$\epsilon_{J/\psi}$\ is the $J/\psi$ binding energy. Such a cross
section leads to a flat tail of $R_{AA}$, and can not explain the
plateau structure in semi-central collisions and the strong
suppression in central collisions observed at
RHIC~\cite{data1,data2}. Generally, the charmonium binding energy
decreases with temperature, and the cross section (\ref{sigma}) is
no longer valid above some dissociation temperature $T_d$, at
which the binding energy becomes zero. In order to take such an
effect into account, we modify the loss term $\alpha\left({\bf
p}_t,{\bf x}_t,\tau|{\bf b}\right)$~\cite{zhuang} by a step
function $\Theta(T_d-T)$,
\begin{equation}
\label{alpha} \alpha_\Psi = {1\over 2E_\Psi}\int{d^3{\bf p}_g\over
(2\pi)^3 2E_g}W_{g\Psi}^{c\bar c}(s)f_g\Theta\left(T-T_c\right)/\
\Theta\left(T_d^\Psi-T\right),
\end{equation}
where $E_{\Psi}$ and $E_g$ are the charmonium and gluon energies,
$W_{g\Psi}^{c\bar c}(s)$ is the transition probability of the
gluon dissociation process as a function of $s=(p_{\Psi}+p_g)^2$,
$f_g\left({\bf p}_g,T,u\right)$ is the gluon thermal distribution,
and $T({\bf x}_t,\tau|{\bf b})$ and $u({\bf x}_t,\tau|{\bf b})$
are the local temperature and velocity of the hot medium. The gain
term $\beta$ can be obtained by the detailed
balance~\cite{zhuang}.

The local temperature $T$ and fluid velocity $u$, which govern the
thermal gluon distribution in $\alpha$ and charm quark
distribution in $\beta$ and the suppression and regeneration
region controlled by the two step functions in (\ref{alpha}), are
determined by hydrodynamic equations. We assume that the produced
partonic plasma reaches local equilibrium at time
$\tau_0=0.6\textrm{ fm}$. After that, the plasma evolves according
to the 2+1 dimensional Bjorken's hydrodynamic eauations,
\begin{eqnarray}
\label{hydro}
 && \partial_{\tau}E+\nabla\cdot{\bf M} = -(E+p)/{\tau}\
,\nonumber\\
&& \partial_{\tau}M_x+\nabla\cdot(M_x{\bf v}) =
-M_x/{\tau}-\partial_xp\ ,
\nonumber\\
&& \partial_{\tau}M_y+\nabla\cdot(M_y{\bf v}) =
-M_y/{\tau}-\partial_yp \
,\nonumber\\
&& \partial_{\tau}R+\nabla\cdot(R{\bf v}) = -R/{\tau}
\end{eqnarray}
with the definitions of $E=(\epsilon+p)\gamma^2-p$, ${\bf
M}=(\epsilon+p)\gamma^2{\bf v}$ and $R=\gamma n$, where $\gamma$
is the Lorentz factor, and $\epsilon, p$ and ${\bf v}$ are the
energy density, pressure and transverse velocity of the QGP. To
close the equations, we take the equation of state of ideal gases
of partons and hadrons with a first order phase transition at
$T_c$. The initial condition for the hydrodynamics at RHIC is the
same as in Ref~\cite{zhuang}. Both the $J/\psi$ direct production
and the feed-down from $\chi_c$\ and $\psi'$ are considered in our
calculation, and the ratio of them in the initial collision is
taken as $6:3:1$~\cite{feeddown}.

In following numerical calculations, we take the dissociation
temperature for $\chi_c$\ and $\psi'$\ as the critical temperature
of the deconfinement phase transition,
$T_d^{\chi_c}=T_d^{\psi'}=T_c$. Since the binding energy of
$J/\psi$ in hot and dense medium is estimated to be less than 220
MeV~\cite{karsch2, rapp2}, we take $\epsilon_{J/\psi}$ = 150 MeV
and the dissociation temperature $T_d^{J/\psi}=1.92T_c$ to fit the
experimental data. From the lattice simulations, the critical
temperature is taken as $T_c$=165 MeV. Extracted from the
experimental data at RHIC energy~\cite{ccpp,ptpp}, the charm quark
and charmonium production cross sections at central rapidity
region in nucleon-nucleon collisions are $d\sigma_{NN}^{c\bar
c}/dy|_{y=0}=120\ \mu$b and
$B_{ll}d\sigma_{NN}^\Psi/dy|_{y=0}$=26.4, 4.4 and 13.2 nb for
$\Psi=J/\psi,\ \psi'$ and $\chi_c$.
%%%%%%%%%%%%%%%%%%%%%%%%%%%%%%%%%%%%%%%%%%%%%%%%%%%%%%%%%%%%%%%%%%%%%
\begin{figure}[!hbt]
 \centering
 \includegraphics[width=0.6\textwidth]{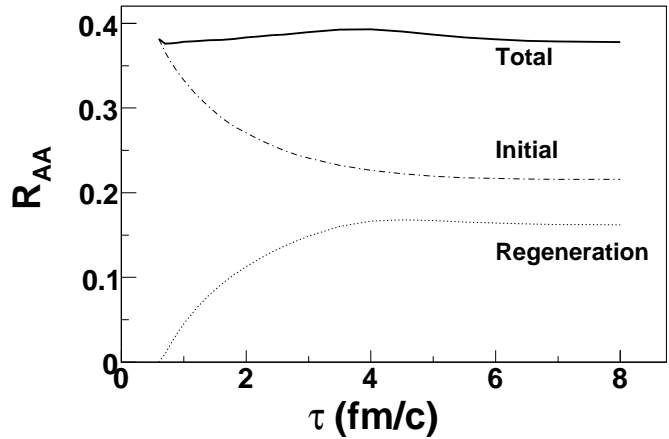}
 \caption{The $J/\psi$ nuclear modification factor as a function of time for central
 Au+Au collisions at RHIC energy. } \label{fig1}
\end{figure}
%%%%%%%%%%%%%%%%%%%%%%%%%%%%%%%%%%%%%%%%%%%%%%%%%%%%%%%%%%%%%%%%%%%%%%

With the known $J/\psi$ distribution $f_{J/\psi}$ as a function of
time and transverse coordinate and momentum, one can easily
extract the space-time evolution of $J/\psi$ production and the
final state distributions determined at the hadronization
hypersurface. The time evolution of the nuclear modification
factor $R_{AA}$ for central Au+Au collisions at RHIC energy is
shown in Fig.\ref{fig1}. The evolution starts at the
thermalization time $\tau_0$. At the beginning, the system is very
hot and both the suppression and regeneration are significant. The
initially produced charmonia, which are treated as initial
condition $f_\Psi({\bf p}_t,{\bf x}_t,\tau_0|{\bf
b})$~\cite{zhuang} of the transport equation (\ref{trans}), suffer
from the strong gluon dissociation and the number drops fast.
Since the regeneration is proportional to the square of charm
quark number but the suppression is linear in gluon number, the
number of the survived regenerated $J/\psi$s increases with time
fast in the beginning. With the expansion of the system, the
temperature decreases, and the suppression and regeneration become
smooth and finally get saturated at the phase transition
temperature $T_c$ which corresponds to the time $\tau\sim 4\
$fm/c. Since we did not consider the suppression and regeneration
in hadronic phase, the initial production and regeneration are
time independent after the phase transition. Due to the
competition between the initial production and regeneration, the
total $J/\psi$ production is almost a constant in the whole
evolution.

\begin{figure}[!hbt]
 \centering
 \includegraphics[width=0.6\textwidth]{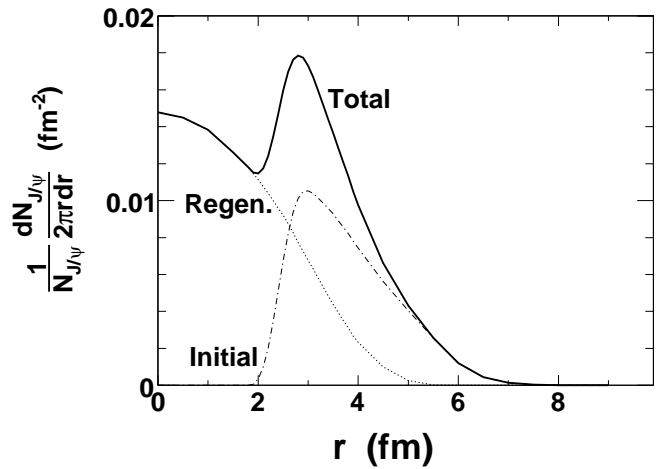}
 \caption{The observed $J/\psi$ number density as a function of the fireball
 radius in the transverse plane for central Au+Au collisions at RHIC energy. }\label{fig2}
\end{figure}

Where are the finally observed $J/\psi$s from, from the central
part of the fireball or the surrounded region? We calculated the
source distribution of the observed $J/\psi$s from the function
$f_{J/\psi}$ in the limit of $\tau\to \infty$. The $J/\psi$ number
density as a function of the fireball radius in the transverse
plane is shown in Fig.\ref{fig2}. The initially produced charmonia
in the central region are fully eaten up by the extremely hot QGP
in the initial stage when the temperature is larger than $T_d$.
Since regeneration happens continuously in the whole QGP region,
those regenerated $J/\psi$s in the later stage with temperature
$T<T_d$ have the probability to survive, and therefore the finally
observed $J/\psi$s coming from the center of the fireball are
mostly regenerated. In the periphery of the fireball where the
temperature is low, both the suppression and regeneration are
weak, and the observed $J/\psi$s from this region are dominated by
the initial production. For those $J/\psi$s coming from the middle
part of the fireball, both the initially produced and regenerated
$J/\psi$s are important.
%%%%%%%%%%%%%%%%%%%%%%%%%%%%%%%%%%%%%%%%%%%%%%%%%%%%%%%%%%%%%%%%%%%%
%%%%%%%%%%%%%%%%%%%%%%%%%%%%%%%%%%%%%%%%%%%%%%%%%%%%%%%%%%%%%%%%%%%%%%

\begin{figure}[!hbt]
\centering
\includegraphics[width=1.0\textwidth]{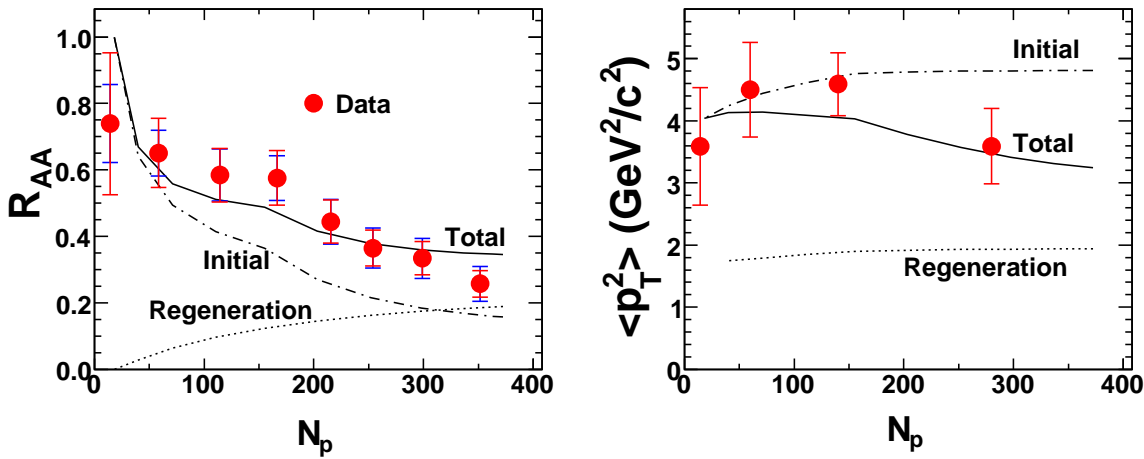}
\vspace{-0.25cm} \caption{The nuclear modification factor $R_{AA}$
(left panel) and averaged transverse momentum square $\langle
p_t^2 \rangle$ (right panel) as functions of participant nucleon
number $N_p$ in mid-rapidity region for Au+Au collisions at RHIC
energy. The data are from PHENIX
collaboration~\cite{data1,data2}.} \label{fig3}
\end{figure}

We now turn to discuss the centrality dependence of the nuclear
modification factor $R_{AA}$ and the averaged transverse momentum
square $\langle p_t^2\rangle$ for $J/\psi$s. The calculated
$R_{AA}$ as a function of the participant nucleon number $N_p$ for
Au+Au collisions at RHIC energy and the comparison with the
experimental data~\cite{data1,data2} are shown in the left panel
of Fig.\ref{fig3}. Since both the suppression and regeneration
increase with centrality, the yield of the initially produced
$J/\psi$s drops monotonously and the yield of the regenerated
$J/\psi$s goes up monotonously. For peripheral and semi-central
collisions where the temperature of the system is low, the
$J/\psi$ production is controlled by the initial production, while
for central collisions where the charm quark density is high, both
the initial production and regeneration are important. From the
data, there exists a flat structure in between $N_p$=\ 60 and 170,
and then the $R_{AA}$ decreases further for $N_p>170$. The plateau
can be explained by the competition between the initial production
and regeneration, and the further decrease in large $N_p$ region
is due to the fact that the maximum temperature of the fireball
created in central collisions is higher than the dissociation
temperature $T_d$. For extremely central collisions around $N_p$=\
350, the difference between the data and the theory is probably
due to the geometry fluctuations which are proved to be important
at SPS energy~\cite{hufner1,fluctuation}.
%%%%%%%%%%%%%%%%%%%%%%%%%%%%%%%%%%%%%%%%%%%%%%%%%%%%%%%%%%%%%%%%%%%%%%%
%%%%%%%%%%%%%%%%%%%%%%%%%%%%%%%%%%%%%%%%%%%%%%%%%%%%%%%%%%%%%%%%%%%%%%%

Different from the global yield, the momentum distribution is more
sensitive to the mechanism of $J/\psi$ production and suppression.
Our theoretical calculation of $J/\psi$ $\langle p_t^2\rangle$ as
a function of centrality for Au+Au collisions at RHIC energy and
the comparison with the experimental data~\cite{data1} are given
in the right panel of Fig.\ref{fig3}. Similar to the calculation
at SPS energy~\cite{hufner1}, the $\langle p_t^2\rangle$ of the
initially produced $J/\psi$s increases with centrality and gets
saturated in central collisions, due to the gluon
multi-scattering\cite{hufner2} with nucleons in the initial state
and the leakage effect which is especially important for high
momentum $J/\psi$s. Since we have assumed kinetic equilibration of
charm quarks in the QGP, the transverse momentum distribution of
the regenerated $J/\psi$s is controlled by the charm quark thermal
distribution, and the corresponding $\langle p_t^2\rangle$ is only
about half of that for initially produced ones. The total $\langle
p_t^2\rangle$ is dominated by the competition between the two
production mechanisms. At high $N_p$ the contribution from the
regeneration becomes important, and the competition leads to a
transverse momentum suppression in central collisions which agrees
well with the data.

We calculated also the $R_{AA}$ and $\langle p_t^2\rangle$ for
Cu-Cu collisions at RHIC energy. Since the colliding energy for
Au+Au and Cu+Cu collisions are the same, and the nuclear geometry
is mainly reflected in the participant nucleon number, the
$R_{AA}$ and $\langle p_t^2\rangle$ for the two kinds of nuclear
collisions are almost the same in the centrality region of
$N_p<110$ where 110 is the maximum $N_p$ for Cu+Cu collisions. Our
calculation agrees with the experimental data~\cite{data3}.

In summary, we have calculated the space-time evolution and the
observed nuclear modification factor and averaged transverse
momentum square of $J/\psi$s produced in relativistic heavy ion
collisions at RHIC energy. While the initial production dominates
the peripheral and semi-central collisions, the initial production
and regeneration are almost equally important for central
collisions.

\section*{Acknowledgements} We are grateful to Xianglei Zhu and
Li Yan for their help in numerical calculations. The work is
supported by the NSFC grant No. 10735040, the 973-project No.
2007CB815000, and the U.S. Department of Energy under Contract No.
DE-AC03-76SF00098.
%--==================================================================

\end{document}